\documentclass[a4paper,11pt]{article}
\usepackage{jcappub} 
\usepackage{lineno}
\usepackage{natbib}
\usepackage{ulem}
\usepackage[T1]{fontenc}

\newcommand\aj{\ref@jnl{AJ}}
\newcommand\psj{\ref@jnl{PSJ}}
\newcommand\araa{\ref@jnl{ARA\&A}}
\newcommand\apj{\ref@jnl{ApJ}}
\newcommand\apjl{\ref@jnl{ApJL}}     
\newcommand\apjs{\ref@jnl{ApJS}}
\newcommand\ao{\ref@jnl{ApOpt}}
\newcommand\apss{\ref@jnl{Ap\&SS}}
\newcommand\aap{\ref@jnl{A\&A}}
\newcommand\aapr{\ref@jnl{A\&A~Rv}}
\newcommand\aaps{\ref@jnl{A\&AS}}
\newcommand\azh{\ref@jnl{AZh}}
\newcommand\baas{\ref@jnl{BAAS}}
\newcommand\icarus{\ref@jnl{Icarus}}
\newcommand\jaavso{\ref@jnl{JAAVSO}}  
\newcommand\jrasc{\ref@jnl{JRASC}}
\newcommand\memras{\ref@jnl{MmRAS}}
\newcommand\mnras{\ref@jnl{MNRAS}}
\newcommand\pra{\ref@jnl{PhRvA}}
\newcommand\prb{\ref@jnl{PhRvB}}
\newcommand\prc{\ref@jnl{PhRvC}}
\newcommand\prd{\ref@jnl{PhRvD}}
\newcommand\pre{\ref@jnl{PhRvE}}
\newcommand\prl{\ref@jnl{PhRvL}}
\newcommand\pasp{\ref@jnl{PASP}}
\newcommand\pasj{\ref@jnl{PASJ}}
\newcommand\qjras{\ref@jnl{QJRAS}}
\newcommand\skytel{\ref@jnl{S\&T}}
\newcommand\solphys{\ref@jnl{SoPh}}
\newcommand\sovast{\ref@jnl{Soviet~Ast.}}
\newcommand\ssr{\ref@jnl{SSRv}}
\newcommand\zap{\ref@jnl{ZA}}
\newcommand\nat{\ref@jnl{Nature}}
\newcommand\iaucirc{\ref@jnl{IAUC}}
\newcommand\aplett{\ref@jnl{Astrophys.~Lett.}}
\newcommand\apspr{\ref@jnl{Astrophys.~Space~Phys.~Res.}}
\newcommand\bain{\ref@jnl{BAN}}
\newcommand\fcp{\ref@jnl{FCPh}}
\newcommand\gca{\ref@jnl{GeoCoA}}
\newcommand\grl{\ref@jnl{Geophys.~Res.~Lett.}}
\newcommand\jcp{\ref@jnl{JChPh}}
\newcommand\jgr{\ref@jnl{J.~Geophys.~Res.}}
\newcommand\jqsrt{\ref@jnl{JQSRT}}
\newcommand\memsai{\ref@jnl{MmSAI}}
\newcommand\nphysa{\ref@jnl{NuPhA}}
\newcommand\physrep{\ref@jnl{PhR}}
\newcommand\physscr{\ref@jnl{PhyS}}
\newcommand\planss{\ref@jnl{Planet.~Space~Sci.}}
\newcommand\procspie{\ref@jnl{Proc.~SPIE}}

\newcommand\actaa{\ref@jnl{AcA}}
\newcommand\caa{\ref@jnl{ChA\&A}}
\newcommand\cjaa{\ref@jnl{ChJA\&A}}
\newcommand\jcap{\ref@jnl{JCAP}}
\newcommand\na{\ref@jnl{NewA}}
\newcommand\nar{\ref@jnl{NewAR}}
\newcommand\pasa{\ref@jnl{PASA}}
\newcommand\rmxaa{\ref@jnl{RMxAA}}

\title{\boldmath Early formation of supermassive black holes from the collapse of strongly self-interacting dark matter}

 \author[a,b]{M. Grant Roberts,}
 \author[a]{Lila Braff,}
 \author[a]{Aarna Garg,}
 \author[a,b]{Stefano Profumo,}
 \author[a,b]{Tesla Jeltema,}
 \author[a,b]{and Jackson O'Donnell}

    \affiliation[a]{Department of Physics, University of California, Santa Cruz (UCSC),
Santa Cruz, CA 95064, USA}
\affiliation[b]{Santa Cruz Institute for Particle Physics (SCIPP),
Santa Cruz, CA 95064, USA}

\emailAdd{migrobert@ucsc.edu}

\abstract{Evidence for high-redshift supermassive black holes challenges standard scenarios for how such objects form in the early universe. Here, we entertain the possibility that a fraction of the cosmological dark matter could be ultra-strongly self interacting. This would imply that gravothermal collapse occur at early times in the cores of dark matter halos, followed by accretion. We study under which conditions on the abundance and interaction strength and structure of such ultra self-interacting dark matter the black holes resulting from the end-point of gravothermal core collapse can seed the observed, early-forming supermassive black holes.  We find, depending on the velocity dependence of the self-interaction cross section, a bimodal structure in the favored parameter space, where data points to either a small collapsing dark matter fraction with a large cross section, or a large fraction and a relatively small cross section. While  self-interaction cross sections with different velocity dependence can explain observations, we find that the best, self-consistent results correspond to a Rutherford-like self-interaction, typical of long-range dark-sector forces with light mediators. We discuss complementary observational probes if this scenario is realized in nature, focusing especially on the expected intermediate mass black holes predicted to exist in smaller galaxies.}

\begin{document}

\newcommand{\mBH}{m_{\text{BH}}}
\newcommand{\mBHseed}{\mBH^{\text{seed}}}
\newcommand{\mBHobs}{\mBH^{\text{obs}}}
\newcommand{\mBHobsi}{m_{\text{BH},i}^{\text{obs}}}
\newcommand{\mBHtheory}{\mBH^{\text{theory}}}
\newcommand{\zcoll}{z_{\text{coll}}}
\newcommand{\zvir}{z_{\text{vir}}}
\newcommand{\zobs}{z_{\text{obs}}}
\newcommand{\cross}{\sigma/m}
\newcommand{\msun}{M_{\odot}}
\newcommand{\tsal}{t_{\text{sal}}}
\newcommand{\trel}{t_{\text{rel}}}
\newcommand{\rhocrit}{\rho_{\text{crit}}}
\newcommand{\cmg}{\text{cm}^{2}\text{g}^{-1}}
\newcommand{\kms}{\text{km}~\text{s}^{-1}}
\newcommand{\angstrom}{\r{A}}
\newcommand\sbullet[1][.5]{\mathbin{\vcenter{\hbox{\scalebox{#1}{$\bullet$}}}}}
\newcommand{\chisq}{\chi^{2}}
\newcommand{\Vmax}{V_{\text{max}}}

\newcommand{\spr}[1]{{\color{red}\bf[SP:  {#1}]}}
\newcommand{\grantcomment}[1]{{\color{blue}\bf[GR:  {#1}]}}

\newcommand{\tesla}[1]{{\color{cyan}\bf[TJ:  {#1}]}}

\maketitle
\flushbottom

\section{Introduction}
\label{sec:intro}

Recent data from the James Webb Space Telescope (JWST) \cite{JWST, jwstlotz} and other observatories~\cite{goulding2023uncover} have unveiled surprisingly early formation epochs for supermassive black holes (SMBHs) \cite{schmidt1963}. SMBHs are black holes (BHs) with masses of $10^{5}~\msun$ and above, thought to be at the center of every galaxy in the universe \cite{ZelNov64, salpeter64, Lynden-Bell}. In general, it is assumed that such objects form and grow through baryonic accretion of gas  \cite{hopkins2010, Volonteri_2010}, and/or through subsequent galaxy mergers \cite{sanders, bhowmick, Volonteri_2010}. However, it is becoming increasingly clear that baryonic processes alone cannot  satisfactorily explain the accelerated growth rate of early-formed SMBHs with redshifts $\gtrsim$ 6-7 
 \cite{Pollack_2015, Choquette_2019}. Quantitatively, high redshift BH seeds would need to grow at a rate that exceeds the Eddington accretion rate \cite{Choquette_2019} -- a putative upper limit on the accretion rate of BHs, where radiative pressure must be balanced by the gravitational force \cite{Salpeter, weinbergeredd, weinberger2}. Super-Eddington accretion is an unlikely process due to radiative and kinetic feedback inhibiting seed growth \cite{Shi_2022, prole2023heavy, Johnson,Milosavljevi__2009,Alvarez_2009,Smith2018}. Therefore, alternative scenarios are warranted; as was first suggested in \cite{2002PhRvL..88j1301B}, here we propose one where the gravothermal collapse of a sub-component of the dark matter (DM) halos forms a seed which eventually grows into the SMBH. 

Previous work on possible explanations for the existence of early-forming SMBHs include the collapse of Population III (Pop III) stars, instabilities of metal-poor gas clouds, and stellar-dynamical methods \cite{Volonteri_2010}. Pop III stars are the first generation of stars in the universe, formed from metal-free gas clouds \cite{Chantavat_2023}. Stars with masses greater than 260$~\msun$ go through accelerated collapse into BHs rather than explosions like supernovae, and can accrete to get to SMBH masses \cite{Bond_1984}. Other possibilities proposed recently include direct collapse triggered by a decaying dark matter component \cite{Lu:2023xoi}, or by the evaporation of late-forming non-stellar BHs \cite{Lu:2024zwa}.

Numerous quasars with $z\gtrsim 6-7$ (roughly 800 Myr after the Big Bang \cite{Feng_2021}) have been observed with estimated masses of several $10^{9}~\msun$, see Refs. \cite{xshooter,xqr,demographics_z_6}, and Ref. \cite{2023ARA&A..61..373F} for a recent review of the overall observational landscape. For a given black hole seed of $10^{5}~\msun$, growing to a supermassive size of $10^{9}~\msun$ through merely baryonic accretion would require a seed of redshift 13.5, or the seed would have to start out more massive than $10^{5}~\msun$, which is presumably possible through stellar or direct collapse processes \cite{Choquette_2019}. Pop III stars — the first generation of stars — form at redshifts between $z = 20 - 30$ \cite{Volonteri_2010} and have brief lifetimes of $3\times10^{6}$ years \cite{Bromm_2003}.  In theory, these stars could collapse soon after their formation, going through Eddington accretion until reaching SMBH mass. Since Pop III stars have light seed masses  ($10^{3} - 10^{4}~\msun$ \cite{Choquette_2019}) and collapse at high redshifts, the resulting black holes  would need to undergo Eddington accretion for hundreds of millions of years to reach SMBH masses. However, during both Eddington and super-Eddington accretion processes, the gas surrounding the seed rises in temperature from radiative feedback. As it heats up, the accretion rate slows down, making it difficult  for Eddington accretion to be sustained over such a prolonged period of time \cite{prole2023heavy, Johnson,Milosavljevi__2009,Alvarez_2009,Smith2018}. Pop III stars may even form very light seeds which could potentially be as small as several tens of $\msun$, meaning a seed would need to undergo an {\it even longer} period of Eddington accretion \cite{prole2023heavy} to reach observed SMBH masses. Furthermore, the masses of the seeds pose an issue for the stability of the environment in which they grow: The seeds must be located in the middle of the gas inflow so that the gas can accrete onto it through Eddington and super-Eddington methods. Since the seeds are light, they often move in random paths around the center, making it problematic to grow through gas accretion \cite{Beckmann_2019,Pfister_2019}. Therefore, it is disfavored to consider Pop III stars as plausible candidates for forming SMBH seeds. 

Mergers also contribute to speeding up BH growth enough to form SMBHs by early-redshifts. Ref.~\cite{bhowmick} conducted simulations for high-redshift SMBHs and found that BHs forming to SMBH sizes during early epochs of the universe are dominated mainly by mergers rather than by accretion. The merger process is, however, largely dependent on the (pre-)existence of heavy seeds \cite{bhowmick}. Furthermore, the type of accretion channel is also critical in how many seeds can grow and merge into SMBHs \cite{bhowmick, jeon, Volonteri_2010}.  The lower the accretion rate, the more mergers are required to get the seed up to SMBH size, or the more continuous the Eddington accretion needs to be \cite{Tanaka_2009}. Models studied in Refs. \cite{bromm2004, yoo2004, rees2006} display that if seeds are positioned optimally by $z\approx$ 6, and velocity kicks are forgone, a series of mergers can indeed produce SMBHs from such seeds \cite{Tanaka_2009}. However, kicks are a crucial effect of mergers \cite{varma}. Though mergers can increase BH growth, the velocity kicks can fully eject BHs from their halos and cease seed growth to SMBHs by the necessary redshifts \cite{yoo2004}.

The collapse of metal-free or metal-poor gas clouds is also a candidate for high-redshift SMBHs \cite{haehnelt, Bromm_2003, loeb1994, Volonteri_2010}.  Metal-poor gas clouds are unstable enough to form BHs rather than fragment and form stars \cite{Trenti_2009, Volonteri_2010}. Angular momentum flow towards the outer edge of the gas clouds results in the central region ($10^{4}~$to $10^{6}~$$\msun$) forming a compact object \cite{Shlosman_Frank_Begelman, Volonteri_2010, Begelman_2006} such as a supermassive star of mass above $5\times 10^{4}~$$\msun$ \cite{Volonteri_2010}. The supermassive star can collapse into Kerr-like BHs, containing 90\% of the stellar mass \cite{Shibata_2002}. In instances where mass accretion is more rapid, gas accumulates on the star. A BH of a few $\msun$ is formed, then grows through baryonic accretion in the envelope, limited by the Eddington accretion rate. The BH is then released from the envelope as a seed which continues to increase its mass via baryonic accretion \cite{Volonteri_2010}. However, supermassive stars can be affected by the rate of mass accumulation. If this rate is too large, the specific entropy of the supermassive star is increased, and the core burns hydrogen,  eventually collapsing into a BH of a few $M_{\odot}$. The feedback flux from the accumulation of matter onto the BH then causes a period of super-Eddington accretion to occur \cite{Volonteri_2010}; however, super-Eddington accretion has several caveats as we explain below.

Another potential way to form early universe SMBHs is through stellar-dynamical methods rather than gaseous processes \cite{Volonteri_2010}. As gases in larger halos start to become enriched from the metals in the early Pop III stars, the fragmentation of these gases can begin to create low mass stars, comparable to the typical  stars that populate our universe today. Once formed, the stars can establish nuclear star clusters \cite{Schneider_2006, Clark_2008}, where it is possible for runaway stellar mergers to occur between massive stars \cite{zwart1999star}.  Mergers of massive stars can create ``very massive stars'' (VMS) which can collapse to BHs of mass $10^{2}~$- $10^{4}~$$\msun$ \cite{Devecchi_2009}.  Star clusters begin forming in protogalactic disks, in which multiple stellar BHs form and merge, potentially forming a BH that can then grow to a SMBH through gaseous Eddington accretion \cite{Kritos}. 
However, at the time of these early universe SMBHs and Pop III stars, the metallicity of the universe is extremely low \cite{Prochaska_2000}. For star formation and fragmentation to be effective at low metallicities, the density of galaxies must cross a metallicity-dependent limit \cite{Devecchi_2009}. This then gives a narrow viable parameter space for star formation, and thereby for VMS formation. Furthermore, as the metallicity of the universe increases, this process becomes less and less efficient \cite{Devecchi_2009}, making stellar-dynamical processes not a likely pathway for early universe SMBH formation. 

None of explanations reviewed above address the rate with which we expect these BHs to grow in order to achieve the masses of observed high redshift SMBHs. Sustained super-Eddington accretion is thought to be a solution to this issue \cite{Pezzulli, Volonteri_2015}. Assuming a ``slim-disk solution,'' which involves radiative-hydrodynamic instabilities, SMBH seeds can grow at about 10 e-folds of intermittent super-Eddington accretion periods \cite{Madau_2014}.  Depending on the environments of the growing seeds and amount of gas present to accrete, super-Eddington accretion can be sustained for more e-folds \cite{Volonteri_2015}. However, there are several caveats that arise with exceeding the Eddington limit. The radiative and kinetic feedback from super-Eddington accretion could accelerate the growth of the BH, but there is also a possibility of BH growth being stunted or ceased due to this feedback \cite{Shi_2022}. Additionally, star formation and fragmentation could occur in the process of BH growth if an excess of material is captured, which also can stop BH growth \cite{Shi_2022}. The BH must be able to capture enough mass for super-Eddington accretion to be effective; otherwise, feedback would only slow growth. Baryonic material needs to be ``gravitationally captured'' \cite{Shi_2022} onto the growing BH, but the rate at which that occurs is unknown. Due to the various difficulties of baryonic formation channels for SMBHs at high redshifts, we propose that the seeds can be formed through the direct collapse of self-interacting DM halos. 

In the current standard model of cosmology, Lambda Cold Dark Matter ($\Lambda$CDM), the Cold Dark Matter (CDM) is postulated to behave like a pressure-less, non-relativistic fluid, supporting in gravitationally collapsed halos a stable, central region  well fit by a Navarro-Frank-White (NFW) density profile \cite{Navarro_1997,plankcollaberation_2014, tran_2024}. CDM falls short, possibly, at explaining some astrophysical phenomena such as the ``core-cusp'' problem \cite{Oman2015, Balberg_2002}, the ``rotation curve diversity'' problem \cite{Oman2015}, and the ``too-big-to-fail'' problem \cite{plankcollaberation_2014}. In light of these potential issues, many studies have considered {\it self-interacting dark matter} (SIDM) \cite{Spergel_2000} (see e.g., \cite{Tulin&Yu_2018,SIDMreview2022} for a review), a possibility that, technically, goes beyond the standard CDM model. SIDM provides a solution to the mentioned astrophysical problems by introducing a non-zero, possibly velocity-dependent, self-interaction cross section among DM particles ($\cross$). 

While $\Lambda$CDM simulations generate ``cuspy'' halos with high densities that steeply increases with radii, observations of dwarf galaxies and low-surface brightness galaxies show a constant density ``core'' \cite{Balberg_2002}‚ a discrepancy known as the ``core-cusp'' problem \cite{Moore_1994,Flores_1994}. As first calculated in Ref. \cite{Balberg_2002}, SIDM resolves this by causing the halo to undergo gravothermal evolution, reducing central densities and forming cores instead \cite{Salucci_2000,Kochanek_2000,Pollack_2015,Yoshida_2000,Dave_2001,Roberts2024}. Varying the relevant SIDM cross sections and velocity-dependence can lead to different core sizes for similarly-sized halos  \cite{Oman2015,Kamada2017,Roberts2024,Kahlhoefer_2019,Spergel_2000,Roberts2024}.

SIDM also explains the ``diversity problem''  where galaxies with similar masses exhibit significantly different rotation curve shapes  \cite{Oman2015,Outmezguine_2023,Ren_2019,Roberts2024}. While $\Lambda$CDM predicts uniformly, steeply rising inner rotation curves due to cuspy profiles \cite{Oman2015}, observations show a diversity of cuspy and cored profiles \cite{Oman2015}. SIDM's self-interactions redistribute heat within the halo, flattening cuspy inner regions and forming a variety of core sizes \cite{Oman2015}, which better aligns with the observed diversity in galaxy rotation curves \cite{Outmezguine_2023,Oman2015,Ren_2019,Roberts2024}. 

The self-interactions of SIDM further offer a solution to the ``too-big-to-fail'' problem \cite{Boylan-Kolchin_2004,Boylan-Kolchin_2011,Pollack_2015,Silverman_2022}. $\Lambda$CDM simulation predict overly dense massive subhalos around galaxies like the Milky Way \cite{Boylan-Kolchin_2011,Boylan_Kolchin_2012}.  These high central densities, would make the halos easily visible as bright satellite galaxies; however, observations show much lower densities \cite{Boylan-Kolchin_2011,Boylan_Kolchin_2012}.  SIDM resolves this by allowing DM particles to scatter with themselves, redistributing energy and forming cores, which lowers central densities, while leaving the outer halo structure largely unchanged \cite{Boylan-Kolchin_2011,Boylan_Kolchin_2012,Pollack_2015,Silverman_2022}.

The self interaction cross section of the DM is, nonetheless, not a free parameter; Rather, it is constrained from direct observations. Fits from strong lensing in clusters imply constraints on $\cross$ within the range $0.1~\cmg$ to $0.5~\cmg$ \cite{Andrade_2021, Sagunski_2021}, and the presence of cores within dwarf and low surface brightness galaxies (LSBs) rotation curves, imply constraints on $\cross$ on the order $1~\cmg$ to $10~\cmg$ when using core expansion type models \cite{Ren_2019,Roberts2024}. However, when allowing for core collapse models, dwarfs and LSBs can instead be well fit with cross sections in the range $20~\cmg$ to $40~\cmg$ \cite{Roberts2024}.

SIDM alone cannot explain black hole formation in the early universe or the rapid growth of SMBHs observed today, except possibly with the aid of baryons \cite{Feng_2021}, or dissipative effects \cite{DissipativeDM}; see \cite{2020ARA&A..58...27I} for additional review. In view of this, Ref.~\cite{Pollack_2015}  introduced the possibility of a subdominant component of {\it ultra self-interacting dark matter} (uSIDM), consisting of a fraction ($f$ $\leq$ 0.1) of the cosmological DM  that is {\it even more strongly interacting} than typical SIDM. In this paper, we generalize and further study the uSIDM model in \cite{Pollack_2015}, by allowing the self-interaction cross section to be velocity-dependent, and by utilizing new observations of high-redshift quasars. Velocity-dependent cross sections can arise via long-range interactions mediated by a gauge boson \cite{Feng_2009,Loeb&Weiner_2011,tulin_2013}, through resonant scattering \cite{Chu_2019, Chu_2020, tsai_2022}, gravitationally bound systems \cite{Braaten_2018}, and particles with strong interactions and significant mass where the cross section increases as the relative velocity decreases, causing scattering to be more frequent at lower velocities \cite{Choi_2017,Hochberg_2015}.

Like SIDM halos, uSIDM halos experience gravothermal evolution \cite{Balberg_2002,Pollack_2015,Outmezguine_2023}. The negative specific heat of the isolated, velocity-dependent uSIDM halo causes the higher density regions in the halo's cuspy center to undergo frequent interactions that spread the mass towards the outer regions. This redistribution of mass and consequently heat, decreases the central density of the halo while its core size increases. Once the core reaches its maximum size the process then reverses; the halo leaves the core-expansion phase and enters the core collapse phase. The core begins to get more and more dense while its size decreases until the core fully collapses into a BH \cite{Pollack_2015,Outmezguine_2023,gadnasr_2023,Balberg_2002,Lynden_1968, Kochanek_2000, Koda_2011, Elbert_2015, Colin_2002,sameie_2018, essig_2019, Kahlhoefer_2019, Zavala_2019, Nishikawa_2020, correa_2021, Turner_2021}.  If this collapse occurs early enough in the halos' evolution, it can provide  seeds for the formation of early SMBHs of adequate mass and redshift \cite{Pollack_2015, Feng_2021,Balberg_2002,gadnasr_2023,Choquette_2019}. This is the centerpiece and focus of the present study.

The structure of the remainder of this paper is as follows: Sec.~\ref{sec:data} details our two choices of quasars samples, as we explain; Sec.~\ref{sec:model} outlines the uSIDM model and the generation of early SMBH seeds; Sec.~\ref{sec:MCMC} describes our statistical approach in comparing the uSIDM framework to data, and shows our results which are then discussed in detail in Sec.~\ref{sec:discussion}. We present our conclusions in Sec.~\ref{sec:conclusion}.

\section{Data and Quasar Samples}\label{sec:data}

We utilize two quasar samples; first, we select objects with two or more independent BH mass and redshift estimates that roughly correspond to similar calibrations. We  queried catalogues X-shooter/ALMA Sample of Quasars in the Epoch of Reionization \cite{xshooter}, XQR-30 \cite{xqr}, and a compilation of samples brought together in \cite{2011ApJ...739...56D} which includes SDSS, CFHQS and SHELLQs high-z quasar surveys, and we found several candidate quasars with redshifts $z \gtrsim 6$.

We initially selected 8 quasars; however, upon initial modeling, we ran into trouble with convergence, in particular for some versions of the velocity dependence of the cross section, and we could not determine if the cause was due to issues with the sample or underlying physics of the uSIDM model. To explore the model and gain insight into the range of allowed physics, we down-selected to a sample of 3 quasars that have the most mass and redshift measurements (two with three measurements of mass and redshift, and one with four measurements). With this sample we were able to get good fits to the data for all versions of the velocity dependence. We refer to this as our {\it 3 quasar sample} (J0100+2802, J0842+1218, P007+04); the corresponding objects are listed in Table~\ref{tab:quasar-data} - with the BH mass (average masses of $1.7\times10^{10}~\msun$, $2.3\times10^{9}~\msun$, $5.2\times10^{9}~\msun$ respectively) and redshift ($z \sim 6.3$, $z \sim 6$, $z \sim 6$ respectively) provided in the listed reference. After this, we were able to extend our analysis to the original 8 quasar sample, which includes 5 quasars with two or more measurements from the referenced catalogues in Table~\ref{tab:8-quasar-data}.

Each mass estimate in Table~\ref{tab:quasar-data} is assumed to have an error of $0.4$ dex due to calibration systematics \cite{Vestergaard:2006xu, shen2011, Marziani:2019don,Mejia-Restrepo:2016let}. The BH masses are derived in the reference papers via the virial theorem and a relation between the radius of the broad line region (BLR) and the 3000-\angstrom~ continuum luminosity ($R_{BLR}$ \&  $L_{\lambda}$ respectively); we outline the derivation here for completeness. Starting with the virial theorem,

\begin{equation}
    \mBH = G^{-1} R_{BLR} V^{2}_{BLR},
\label{eq:bh-virial-theorem}
\end{equation}

\noindent where $V_{BLR}$ is the circular velocity of the line-emitting gas at $R_{BLR}$. We assume a power law dependence for the $R_{BLR}-L_{\lambda}$ relation:

\begin{equation}
    R_{BLR} = A\left(\frac{L_{\lambda}}{L_{\gamma}}\right)^{B},
\label{eq:R_BLR}
\end{equation}

\noindent where $A$, $B$, and $L_{\gamma}$ are constants to be fit. The circular velocity $V_{BLR}$ is related to the full width at half maximum (FWHM) of the emission line via $V_{BLR} = f_{BLR} \text{FWHM}$ where $f_{BLR}$ is a geometric factor; though in most cases, setting $f_{BLR} = 1$ is a good approximation \cite{McLure_Dunlop2004}. Combining Eq.~\ref{eq:bh-virial-theorem} and Eq.~\ref{eq:R_BLR} with $V_{BLR}$ and rescaling, we can derive the BH mass estimator:

\begin{equation}
    \frac{\mBH}{\msun} = 10^{a}\left(\frac{L_{\lambda}}{10^{44}~\text{erg}~\text{s}^{-1}}\right)^{b} \left(\frac{\text{FWHM}}{\kms}\right)^{2},
\label{eq:bh-mass-estimator}
\end{equation}

\noindent where $a$ and $b$ are constants to be fit. Various values for these constants (or analogues thereof) have been proposed in the literature, see e.g. Ref.~\cite{shen2011, Vestergaard:2006xu, Mejia-Restrepo:2016let, Marziani:2019don}. The values of $a$ and $b$ depend for instance on which spectral line is analyzed (e.g. Mg II or C IV), and how the calibration is done. Typical values for $a$ and $b$ range from 0.5 to 0.9 and 0.5 to 0.62, respectively. The error inferred from Eq.~\ref{eq:bh-mass-estimator} is in the range of 0.3-0.5 dex, which typically dominates measurement errors from the spectral lines themselves. For our sample, we adopt an error value of 0.4 dex on each BH mass estimate and we have chosen BH mass estimates from fits to the Mg II lines that correspond to values of $a$ and $b$ fit in Ref.~\cite{shen2011}.

\begin{table}[t]
\centering

\begin{tabular}{|c|c|c|c|}
\hline

 Quasar &  $\log_{10}M_{\sbullet}~(M_{\odot})$ & z & Reference\\\hline
J0100+2802 & 10.29 & 6.327 & \cite{xshooter}\\
& 10.09 & 6.316  & \cite{xqr}\\
& 10.33 & 6.300 & \cite{demographics_z_6}\\\hline
J0842+1218 & 9.404 & 6.075 & \cite{xshooter}\\
& 9.300 & 6.067 & \cite{xqr}\\
& 9.520 & 6.070 & \cite{demographics_z_6}\\
& 9.230 & 6.069 & \cite{2011ApJ...739...56D}\\\hline
P007+04 & 9.356 & 6.002  & \cite{xshooter}\\
& 9.890 & 5.954  & \cite{xqr}\\
& 9.910 & 5.980  & \cite{demographics_z_6}\\\hline
\end{tabular}

\caption{Here we list the measurements for the masses and redshifts of each quasar in our sample. We also list the corresponding reference where each measurement was calculated. All of the BH mass values were derived using mass estimators based off of Mg II lines (see e.g., \cite{shen2011}). Different choices of the fit parameters for the BH mass estimate exist but primarily our sources use the fit parameters from \cite{shen2011}. There are measurement errors associated with each spectral line; however, this error is superseded by the error in the mass estimate relation, which can be between 0.3-0.5 dex. For our masses, we chose to use a uniform error value of 0.4 dex across the sample (see discussion in the main text). 
}
\label{tab:quasar-data}
\end{table}

\section{Black Hole Seeds from Gravothermal Collapse with ultra-Self-Interacting Dark Matter}\label{sec:model}

We describe here in detail the estimate for the formation of the SMBH seed from the gravothermal collapse of the uSIDM component under investigation. We adapt the methodology of Ref.~\cite{Pollack_2015} as follows: first, we compute the collapse redshift, $\zcoll$, for forming the BH seed from  gravothermal collapse of the halo. This redshift can be determined by solving,

\begin{equation}
    t(\zcoll) - t(\zvir) = 455.65 ~\trel\left(f, \cross, m_{200}, \zvir, c_{200}\right),
\label{eq:zcoll-constraint}
\end{equation}

\noindent where the relax time, $t_{\rm rel}$, is defined as

\begin{multline}
    \trel\left(f, \cross, m_{200}, z, c_{200}\right) = 0.354 ~\text{Myr} \left(\frac{m_{200}}{10^{12}\msun}\right)^{-1/3} \left(\frac{k_{c}(c_{200})}{k_{c}(9)}\right)^{3/2} \left(\frac{c_{200}}{9}\right)^{-7/2} \\  \times \left(\frac{\rhocrit(z)}{\rhocrit(z=15)}\right)^{-7/6}\left(\frac{f \cross}{1 ~ \cmg}\right)^{-1},
\label{eq:trel}
\end{multline}

\noindent with $k_{c}(x) = \ln(1 + x) - x/(1 + x)$ and $\rhocrit(z)$  the critical density of the universe at a given redshift. The time integral is defined in the usual way,

\begin{equation}
    t(z) = \int_{z}^{\infty}\frac{dz'}{(1+z')H(z')},
\label{eq:cosmo-time}
\end{equation}

\noindent with $H(z)$‚ the Hubble rate‚ as a function of redshift (Note that we use here the  cosmological parameters from \cite{Planck2015} to estimate $H(z)$). In principle one must numerically solve Eq.~(\ref{eq:zcoll-constraint}), but under the approximation that $\Omega_{\text{rad}} \sim 0$ and that $(1+z)^{3}~\Omega_{m} >> \Omega_{\Lambda}$ (which is true for $z \gtrsim 0.3$),  Eq.~(\ref{eq:cosmo-time}) can be solved analytically, which, in turn, also allows Eq.~(\ref{eq:zcoll-constraint}) to be solved analytically for $\zvir$. 

Given the collapse redshift, we use Eq. (31) of Ref.~\cite{Pollack_2015} in order to estimate  the BH seed mass, $\mBHseed$:

\begin{equation}
    \frac{\mBH(z)}{m_{200}} \simeq \frac{0.025 f}{\ln{(c_{200})} - \frac{c_{200}}{1 + c_{200}}}.
\label{eq:mBH-calc}
\end{equation}
We note that the factor 2.5\% in the equation above reflects what is found in Eq.~(31) of \cite{Pollack_2015}. The relevant universal relations do not depend on the underlying SIDM cross-section, up to the point of the transition to the short mean free path (smfp) regime, but at this point the relativistic instability will end in collapse regardless - the only difference being the timescale. However, because this is in the super exponential part of the density evolution, the differences in timescales here are extremely small and will presumably not matter much. The equation above also implies that the mass that goes into the BH is approximately 2.5\% of $m_{200}$ of the halo, as can be seen e.g. in figure 4 of Ref.~\cite{Pollack_2015} Note that the dynamical equations are integrated up to approximately 450 $t_{\text{rel}}$ and have reached the smfp regime in all cases.

After seed formation, the BH mass grows, due to accretion, through the observation redshift $\zobs$, to a value that should be compared with the observed BH mass, $\mBHobs$ at that redshift $\zobs$. Growth via accretion can be parameterized by the number of e-folds ($N_{e}$) during which the SMBH grows, as:

\begin{equation}
    t(\zobs) = t(\zcoll) + N_{e}\tsal,
\label{eq:e-folds-constraint}
\end{equation}

\noindent where $\tsal$ is the Salpeter time \cite{Salpeter} (the time scale at which it would take a BH to double its mass via Eddington accretion), which, in turn, depends on the radiative efficiency $\epsilon_{r}$ and the Thompson cross section $\sigma_{T}$:

\begin{equation}
    \tsal = \frac{\epsilon_{r}\sigma_{T}c}{4\pi G m_{p}} \approx \left(\frac{\epsilon_{r}}{0.1}\right)45.1~\text{Myr}.
\label{eq:salpeter-time}
\end{equation}

\noindent Following \cite{Pollack_2015} we set $\epsilon_{r} = 0.1$. The seed then grows exponentially with the number of e-folds,

\begin{equation}
    \mBHtheory(\zobs) = \mBHseed(\zcoll)\exp({N_{e}}).
\label{eq:seed-growth}
\end{equation}

The model outlined above depends on several input parameters: \{$\zvir$, $f$, $\cross$, $m_{200}$, $c_{200}$\}. We  eliminate $\zvir$ as a separate degree of freedom  by setting $\zvir \sim 13.5$, motivated by the fact that after initial MCMC testing, we found that the results are largely insensitive to the choice of $\zvir$; in addition, Ref.~\cite{Pollack_2015} finds that the bound on redshift formation is $z \gtrsim$ 13.5. The combination of the value $f\cross$, based on the analysis from \cite{Pollack_2015}, should take on values $O(1)~\cmg$ in order to be consistent with the optical thickness constraint. The latter amounts to the requirement that the dynamical timescale to collapse be much shorter than the relaxation time, which, in turn implies that 
\begin{equation}
f\cross\le \frac{1}{r_s\rho_s}\sim O(1)\ \cmg,  
\end{equation}
with $r_s$ and $\rho_s$ the usual characteristic scale radius and scale density of a Navarro-Frenk-White dark matter density profile \cite{Navarro_1997}. What this implies for the seed formation is that {\it we generically expect two degenerate solutions within the uSIDM parameter space}. Because the seed mass prediction, Eq.~(\ref{eq:mBH-calc}) is independent of the cross section, and $\trel$ only depends on the combination $f\cross$, and not on their individual values, the two generic predictions for the evolution trajectory of the BH seed we expect are as follows: the seed can either be 

(1)  formed with a smaller mass and have large subsequent accretion growth ($f$ small and $N_{e}$ large), or 

(2) it can be formed with a larger mass and undergo less accretion growth ($f$ large and $N_{e}$ small). 

Within the uSIDM model alone there is no way to distinguish between the two evolution trajectories and so the distinction between the two cases must be driven and selected by data. We also expect that one or the other solution be favored by different velocity dependence structures, which we discuss next.

Let $\sigma_{0}/m$ be the uSIDM velocity-independent cross section; we can then assume the velocity-dependent structure of the uSIDM cross section ($\sigma(v)/m$) to follow a power law in velocity:

\begin{equation}
    \sigma(v)/m = \sigma_{0}/m \left(\frac{v}{v_{\text{ref}}}\right)^{n},
\end{equation}

\noindent where $v_{\text{ref}}$ is a reference velocity. Assuming virialization, we can convert a velocity to a mass via $v = \sqrt{\frac{G m_{200}}{r_{200}}}$; there $r_{200}\sim m_{200}^{1/3}$, thus $v\sim m_{200}^{1/3}$. Hereafter, we pick a reference velocity such that it translates to a reference mass $m_{\text{ref}} \sim 10^{12} M_{\odot}$. The velocity dependent cross section then becomes:

\begin{equation}
    \sigma(v)/m = \sigma_{0}/m \left(\frac{m_{200}}{10^{12}\ m_{\odot}}\right)^{n/3}.
\end{equation}

\noindent In order to translate from uSIDM into SIDM, we use Eq. (20) in \cite{Pollack_2015}, which implies

\begin{equation}
    \left(\frac{\sigma(v)}{m}\right)^{\text{SIDM}} = f \left(\frac{\sigma(v)}{m}\right)^{\text{uSIDM}}.
\end{equation}

\noindent Note that if $f = 1$ and $n = 0$ (i.e., no velocity dependence), then i $\sigma_{0}/m = \sigma(v)/m = \sigma(v)^{\text{SIDM}}/m$. But if $f = 1$ and $n \neq 0$, then $ \sigma_{0}/m \neq \sigma(v)/m$, but $\sigma(v)/m$ still is equal to $\sigma(v)^{\text{SIDM}}/m$. 

In the following analysis we consider four fiducial values of the power law dependence, $n = +1$, $0$, $-2$, and $-4$. These are motivated by phenomenological and model-building considerations: for instance,
$n=-1$ would arise from a Yukawa potential, i.e. the case of a massive boson mediator , $n=-2$ occurs for an interaction structure typical of electric
dipole moments, and $n=-4$ for a millicharged or a darkly-charged or long-range dark force \cite{Tulin:2017ara}.

\section{MCMC details and implementations}\label{sec:MCMC}

We test the model under consideration, which as mentioned takes as inputs the parameters \{$f$, $\cross$, $m_{200}$\}, utilizing an MCMC algorithm with the data described in Sec.~\ref{sec:data} above. To calculate concentrations, we use the concentration-mass relation from \cite{Diemer2019}, and allow a 0.33 dex scatter (about 3$\sigma$) around the median. Specifically we generate a $c_{200}$ from the input $m_{200}$ and then sample the median with the 0.33 dex scatter. We make use of the sampler ~\texttt{emcee}~\cite{emcee} to run the MCMC, and we heavily use the package~\texttt{Colossus}~\cite{colossus} for our cosmology and concentration-mass relation.  

In order to extract the uSIDM cross section and abundance, we utilize a hierarchical model which  allows us to keep the same $f$ and $\cross$ for every quasar, while allowing a separate $m_{200}$ for each. In this way, for example, in the 3 quasar sample have 5 total MCMC parameters: 3 different $m_{200}$ parameters, 1 parameter for $f$, and 1 parameter for $\cross$. We use a $\chisq$ for our likelihood function, of the form:

\begin{equation}
    \chisq = \sum_{i=1}^{N_{\text{quasar}}}\frac{\left(\log_{10}\text{Avg}(\mBHobsi)-\log_{10}\mBHtheory(\text{Avg}(z_{\text{obs},i}))\right)^{2}}{(\sigma_{m}/\sqrt{N_{m,i}})^{2}},
\end{equation}

\noindent where, for each quasar, we evaluate the model at the average of the observation redshifts, $\log_{10}\mBHtheory(\text{Avg}(z_{\text{obs},i}))$, and compare to the weighted average of the measurements of the BH's estimated mass, $\log_{10}\text{Avg}(\mBHobsi)$. We accordingly scale the error in the mass estimate, $\sigma_{m}$, by the square root of the number of mass estimates for each quasar, $N_{m}$, in the standard way.
We use a flat prior on the logarithm of each of the parameters. For the masses, we set the boundary range from $\log_{10}m_{200} = (7.0, 15.5)~\msun$ with the exception of the second mass parameter, as the quasar it corresponds to is the most massive quasar in the sample by an order of magnitude, so we adjust the upper limit of the prior range to $16.5~\msun$. The bounds on the uSIDM fraction are $\log_{10}f = (-10, 0)$, and the bounds on the uSIDM cross section are $\log_{10}\cross = (0, 10)~\cmg$. We also impose two additional priors: 

1. $f\cross < 10~\cmg$ which insures that the model stays approximately within the optical thickness bounds, as discussed above, while also allowing for room to explore scatter in the concentration-mass relation, and,

2. $N_{e} > 0$ which enforces that {\it some} accretion growth occur. 

\noindent Corner plots showing the results of our MCMC are presented in Fig.~\ref{fig:MCMC_params}-\ref{fig:derived_params_n=-4} for the 3 quasar sample and in Fig.~\ref{fig:MCMC_params_8_quasar} for the 8 quasar sample. We discuss the results we obtain in the following section.

\section{Discussion}\label{sec:discussion}

The first key result of the present study is that if a small, ${\cal O}(10^{-4}-10^{-3})$ fraction of the dark matter is ultra-strongly self interacting, gravothermal collapse produces seeds that can successfully explain recent observations of very-high redshift supermassive black holes. Specifically, this scenario leads to two solutions: Fig.~\ref{fig:MCMC_params} illustrates, for instance,  how our fit to the 3 quasar sample with a cross section featuring a velocity dependence $\sigma(v)\sim \sigma_0 v^{-2}$, i.e., in our notation, $n = -2$, prefers two  modes: a low uSIDM fraction with a correspondingly larger cross section ($f \sim 10^{-8}$, $\cross \sim 10^{8.5}~\cmg$) and, vice versa, a larger uSIDM fraction  and a lower cross section ($f \sim 10^{-2}$, $\cross \sim 10^{2.5}~\cmg$), still compatible, however, with SIDM limits. This bimodality is also reflected in the halo mass distributions, since the uSIDM fraction and halo masses are inversely related to each other, and both are correlated to the predicted BH seed mass via Eq.~\ref{eq:mBH-calc}.

We find, however, that the bimodality is {\it not} reflected in the inferred $\mBHseed$ distribution, as shown in Fig.~\ref{fig:derived_params} (except possibly a tail in the inferred distribution). Rather, the final predicted BH mass, $\mBHtheory$ at $\zobs$ for each quasar, is very close to a Gaussian. An unexpected finding is, however, the absence of bimodality in the number of e-folds. While physically we expect that a seed formed from a smaller uSIDM fraction would need more accretion to match the observed BH mass (and vice versa), what we actually find is that the number of e-folds is fairly sharply peaked at $N_{e}\approx 10-12$ in either case, meaning that both for small and large uSIDM fractions accretion needs to be significant. In fact, the only significant driver for the number of accretion e-folds is the halo mass: in order to create a roughly constant mass seed with roughly the same number of e-folds of accretion, the halo mass must necessarily act opposite to the uSIDM fraction (see Eq.~\ref{eq:mBH-calc}) — which is what is observed in the resulting halo mass distributions in Fig.~\ref{fig:MCMC_params}. Because of this, the halo mass can become so large as to be unphysical - which we find to be the case for the $n = 0$, $n = 1$, and the low $f$, high cross section, $n = -2$ solutions. 

The results for the  $n=-4$ case are shown in Fig.~\ref{fig:MCMC_params_n=-4} and \ref{fig:derived_params_n=-4}. The first figure shows that, unlike in the $n=-2$ case, the $n=-4$ case does not show a bimodal structure in the distributions for $f$ and $\cross$: the MCMC converges smoothly to a sharply peaked distribution in both, with a slight, and expected, degeneracy. The median values of the distributions are $f \sim 10^{-3.5}$ and $\cross \sim 10^{3.82}~\cmg$; which are more in line with the small fraction $n = -2$ solution, though the overall distributions for $n = -4$ and $n = -2$ do have significant overlap. The seed masses and accretion rates shown in Fig.~\ref{fig:derived_params_n=-4}; in comparison to the $n=-2$ case, these show more pronounced tails for lower accretion rates and larger seed masses. 

Note that we do not show the other two power-law corner plots ($n = 0$ and $n = 1$) as they do not display the dual modes with the same significant statistics as the $n = -2$ model does. However, they also do result in very similar values of the derived parameters as in Fig.~\ref{fig:derived_params}. 

Fig.~\ref{fig:model_projection} shows the black hole mass as a function of the dark matter halo mass, for a uSIDM fraction that is fixed to the median value from each velocity dependent model. Each line corresponds to a different number of e-folds of accretion, with 1 and 2 $\sigma$ contours in $f$ placed on the $N_{e} = 10$ line. For comparison, we also show constraints on BH masses in low-redshift nuclear star clusters \cite{Neumayer_2020} and dwarf galaxies \cite{Reines_2013,Baldassare_2015}.
We point out in particular the recent lower limit for the mass of a putative IMBH at the center of $\omega$-Centauri, estimated to have a mass of approximately 8,200 $\msun$ \cite{Haberle_2024}. Fig.~\ref{fig:model_projection} shows that the uSIDM model under consideration can in fact explain at least a sub-population of SMBHs as well as potential IMBH candidates (for examples of other potential IMBH candidates see e.g. \cite{Mezcua_2017, Pasham_2014, K_z_ltan_2017}).

The figure also shows median values of the halo masses and BH masses for each of the models as green stars (and blue crosses in the case of the small fraction $n = -2$ solution). While the $n = 0$ and $n = 1$ models have median $f$ values that give rise to error bars covering the entire sample of data, the $m_{200}$ corresponding to each of the halos are much too large for a single galaxy, as $m_{200}\sim 10^{14} \msun -10^{15} \msun$ are galaxy cluster scale masses. Thus, we can rule out the $n = 0$ and $n = 1$ models since they prefer unphysical halo masses.

The $n = -2$ model solution with the small uSIDM fraction presents a similar issue: the 1 and 2 $\sigma$ contours cover nearly all of the points but imply halo masses that are unreasonably large. We note that the large uSIDM fraction for the $n = -2$ solution has, instead, reasonable halo masses; however, this model cannot simultaneously explain observed low-redshift BHs in dwarfs and star clusters unless the accretion in these systems is lower than in the high-redshift quasars. The $n=-4$ case can fit both the high and low-redshift data within $\sim3 \sigma$ on the uSIDM fraction. We thus conclude that the favored model is the $n = -4$ power law which implies both reasonable halo masses as well as can explain most of the data points. 

Fig.~\ref{fig:MCMC_params_8_quasar} (left), shows the corner plots for $f$ and $\cross$ corresponding to the full 8 quasar sample (which we detail on in table \ref{tab:8-quasar-data}), for the $n = -2$ power law. The inclusion of a large number of quasars narrows the parameter space and forces the dual solutions to disappear: the lower $f$ and large $\cross$ solution is now disfavored, with the higher $f$ solution favored, instead, by data. Interestingly, for the $n = -4$ power law, the parameter distributions are similar when we fit with either the 3 or 8 quasar sample, see Fig.~\ref{fig:MCMC_params_8_quasar} (right), but are a bit  narrower - indicating that the 8 quasar sample has, as expected, more constraining power. We find that neither the $n = 0$ nor $n = 1$ power laws provide satisfactory fits for the full 8 quasar sample. This reinforces our conclusion from Fig.~\ref{fig:model_projection} that these power law dependencies are incompatible with current data. 

We note that uSIDM collapse happens concurrently with the formation of first-generation stars (Pop III). Whether or not Pop III star formation is accelerated, or frustrated by processes related to uSIDM collapse is unknown, and would need dedicated simulations. We leave this task to future work.

\begin{table}[t]
\centering

\begin{tabular}{|c|c|c|c|}
\hline

 Quasar &  $\log_{10}M_{\sbullet}~(M_{\odot})$ & z & Reference\\\hline
J0100+2802 & 10.29 & 6.327 & \cite{xshooter}\\
& 10.09 & 6.316  & \cite{xqr}\\
& 10.33 & 6.300 & \cite{demographics_z_6}\\\hline
J0842+1218 & 9.404 & 6.075 & \cite{xshooter}\\
& 9.300 & 6.067 & \cite{xqr}\\
& 9.520 & 6.070 & \cite{demographics_z_6}\\
& 9.230 & 6.069 & \cite{2011ApJ...739...56D}\\\hline
P007+04 & 9.356 & 6.002  & \cite{xshooter}\\
& 9.890 & 5.954  & \cite{xqr}\\
& 9.910 & 5.980  & \cite{demographics_z_6}\\\hline
P036+03 & 9.775 & 6.541 & \cite{xshooter}\\
& 9.430 & 6.527 & \cite{xqr}\\\hline
P183+05 & 9.674 & 6.439 & \cite{xshooter}\\
& 9.410 & 6.428  & \cite{xqr}\\\hline
P231-20 & 9.806 & 6.587 & \cite{xshooter}\\
& 9.520 & 6.564 & \cite{xqr}\\\hline
P323+12 & 9.243 & 6.587 & \cite{xshooter}\\
& 8.920	& 6.000 & \cite{xqr}\\\hline
P359-06 & 9.577 & 6.172 & \cite{xshooter}\\
& 9.000 & 6.169 & \cite{xqr}\\\hline
\end{tabular}

\caption{The full 8 quasar sample, with the same notation of Table~\ref{tab:quasar-data}.}
\label{tab:8-quasar-data}
\end{table}

\section{Conclusions}\label{sec:conclusion} 

Supermassive black holes are observed to form much earlier than previously thought. Observations are broadly at odds with a standard formation pathway that starts with black holes resulting from the collapse of Pop III stars and evolve via a combination of mergers and accretion. Several solutions to this conundrum have emerged, including some advocating for new physics, and some developing caveats within the standard scenario.

In the present study, we focused on a scenario involving a sub-sector of the cosmological dark matter (consisting of a fraction of a per mille of the total dark matter mass density) that is {\it ultra}-self-interacting. While only a small fraction of the dark matter, such a sector can trigger the gravothermal collapse of the core of dark matter halos into black holes at early times. In turn, with standard accretion rates, such black hole ``seeds'' can produce the observed, high-redshift supermassive black holes.

\begin{figure}[h]
\includegraphics[scale=0.5]{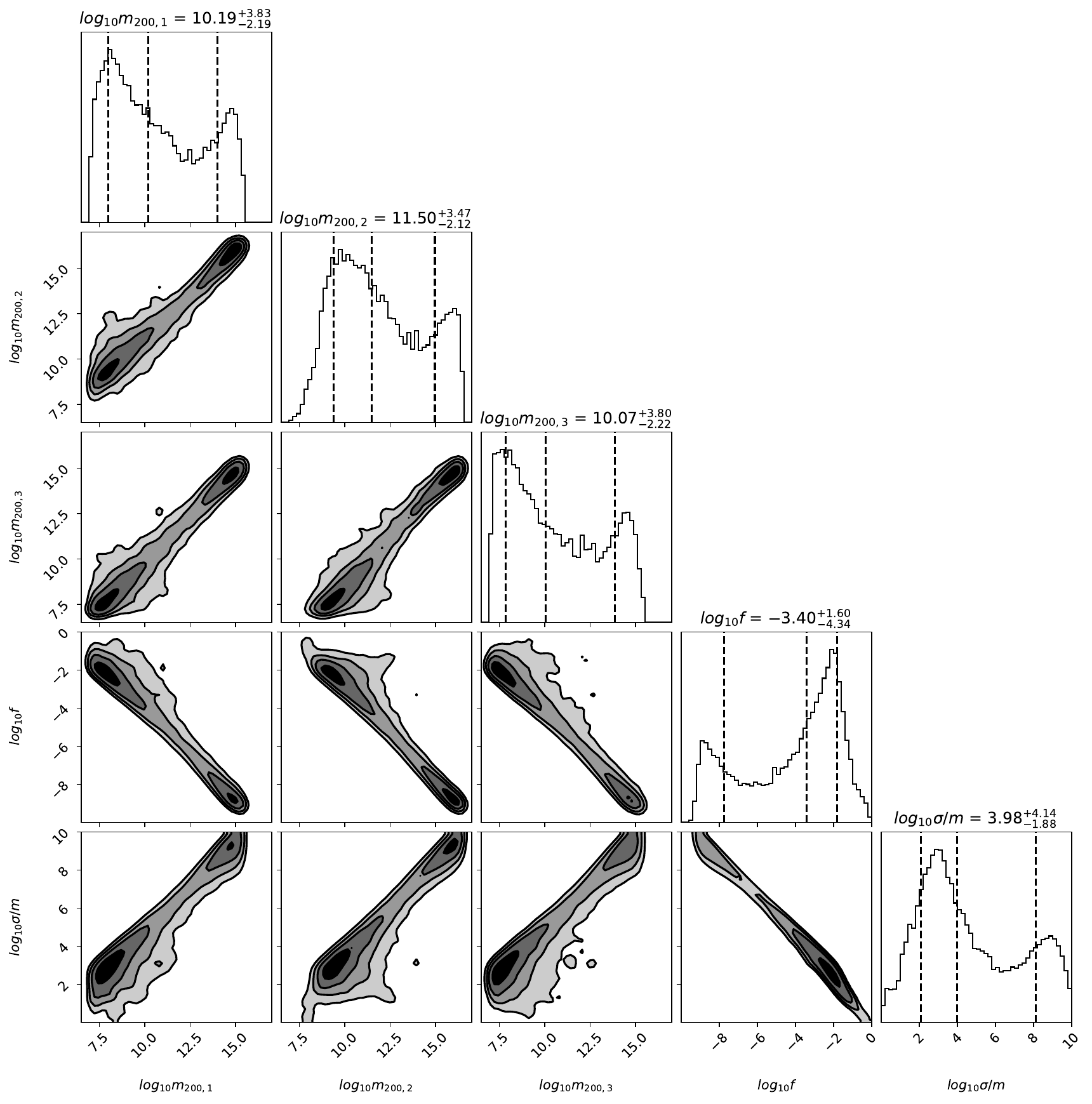}
\caption{We show the MCMC corner plot for each of the DM halo masses ($m_{200}$ in $M_{\odot}$) as well as the uSIDM fraction ($f$) and uSIDM cross section ($\cross$ in $\cmg$) for the $n = -2$ velocity power law model.}
\label{fig:MCMC_params}
\end{figure}

\begin{figure}[h]
\includegraphics[scale=0.35]{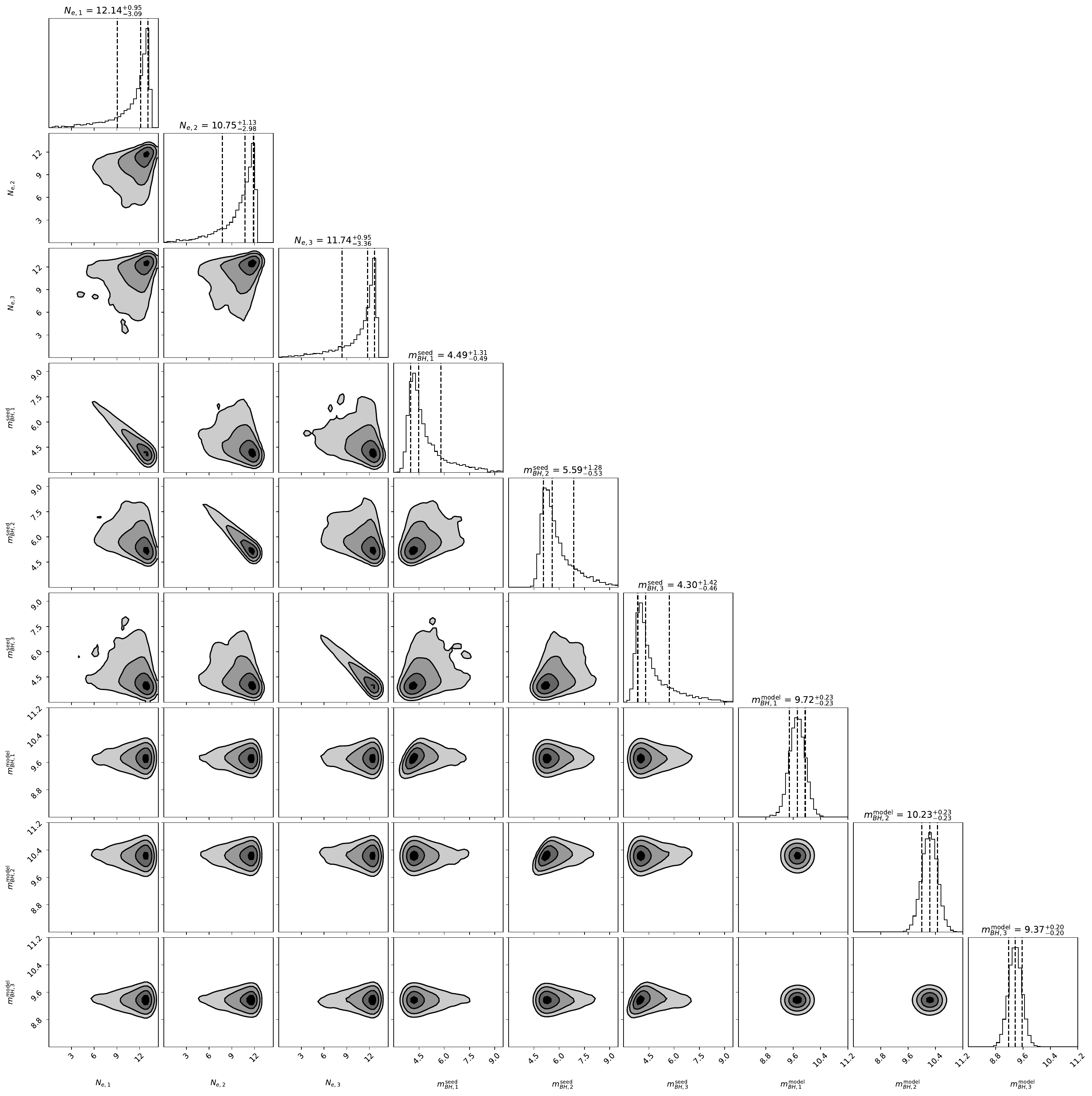}
\caption{Here we show the derived parameter distributions that correspond to the parameters in Fig.~\ref{fig:MCMC_params} for each of the e-fold numbers $N_{e}$, seed BH masses $\mBHseed$, and predicted observed BH masses $\mBHtheory$ for the $n = -2$ power law.}
\label{fig:derived_params}
\end{figure}

\begin{figure}[h]
\includegraphics[scale=0.5]{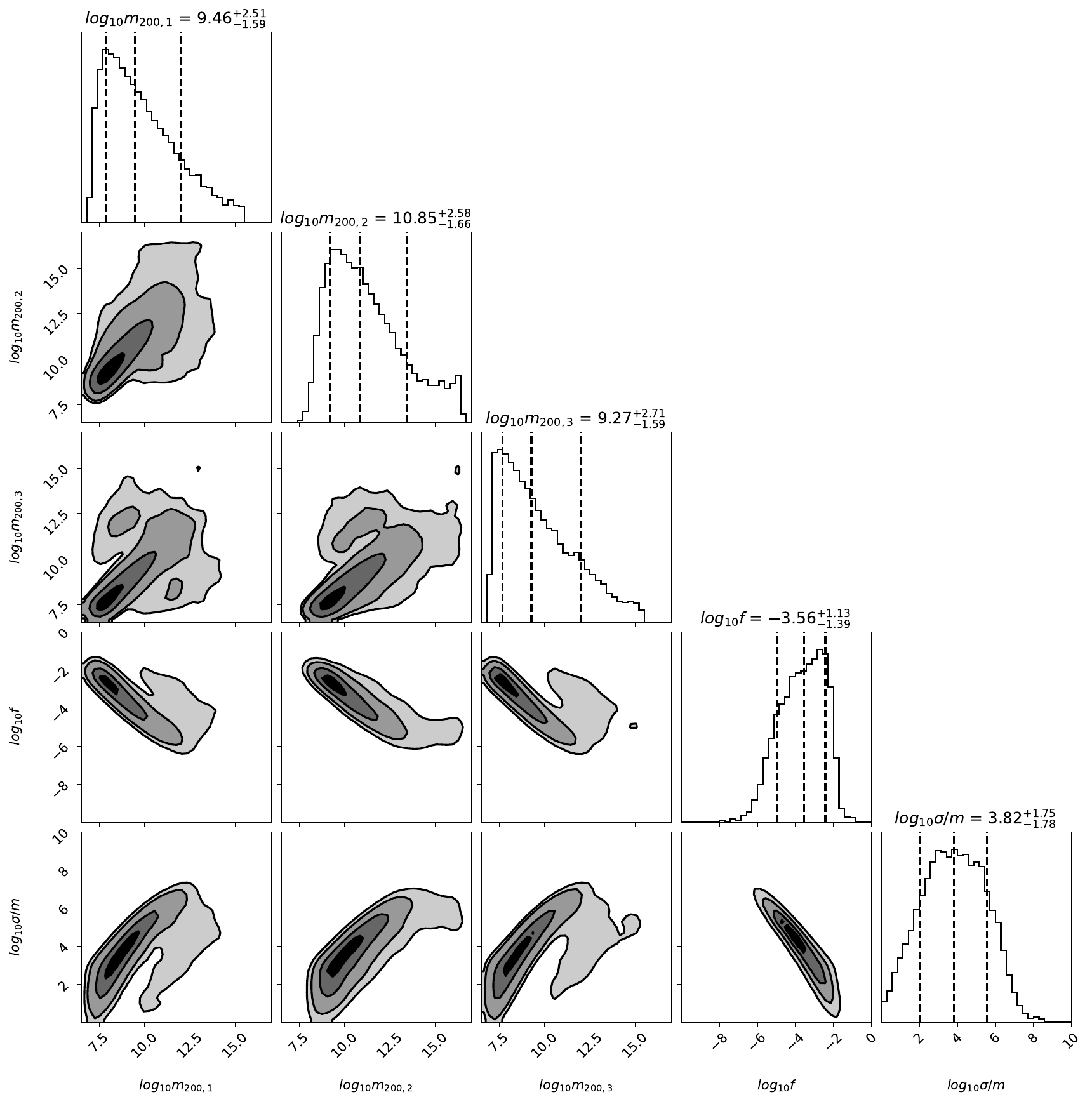}
\caption{We show the MCMC corner plot for each of the DM halo masses ($m_{200}$ in $M_{\odot}$) as well as the uSIDM fraction ($f$) and uSIDM cross section ($\cross$ in $\cmg$) for the $n = -4$ velocity power law model.}
\label{fig:MCMC_params_n=-4}
\end{figure}

\begin{figure}[h]
\includegraphics[scale=0.35]{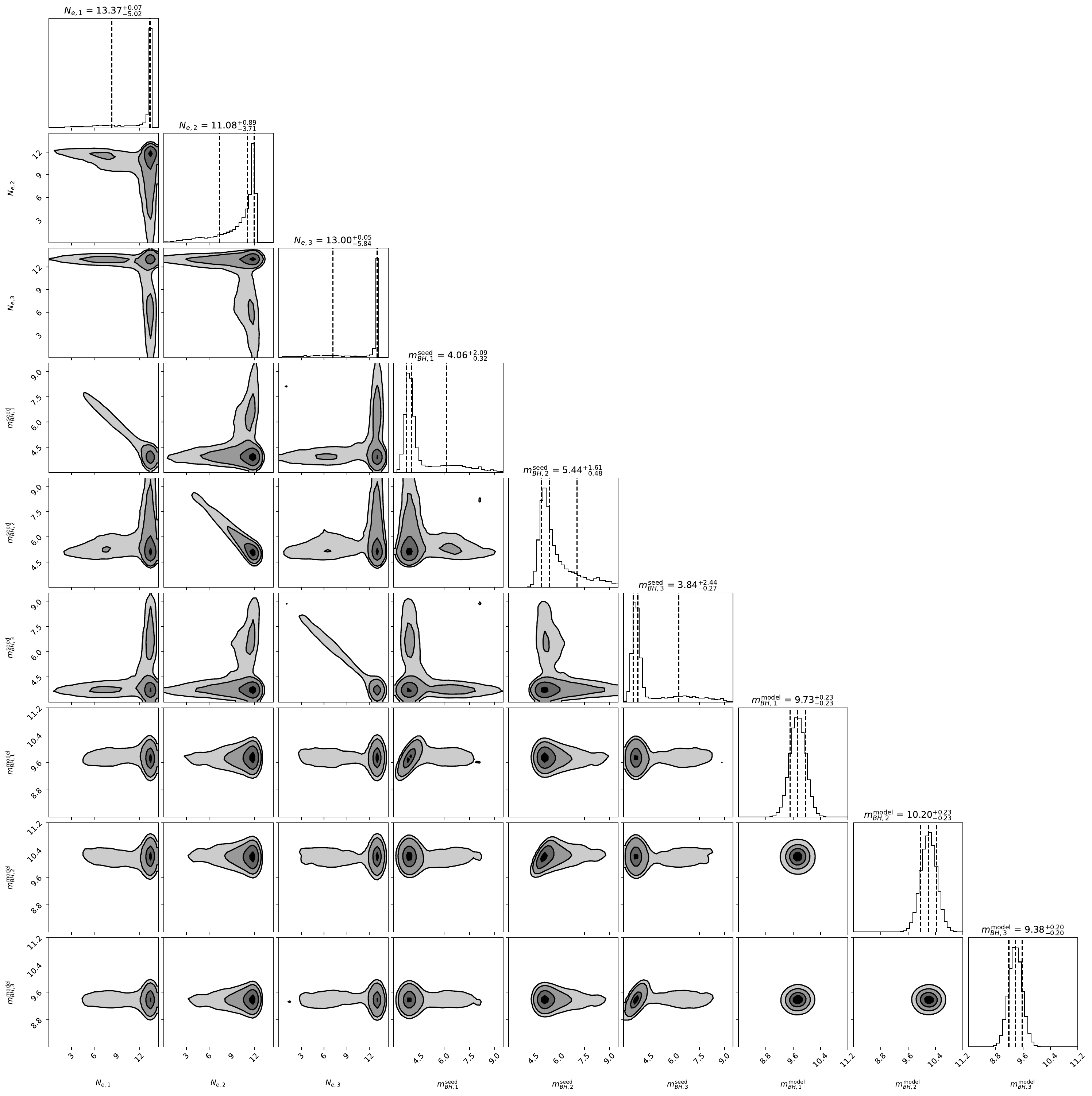}
\caption{Here we show the derived parameter distributions that correspond to the parameters in Fig.~\ref{fig:MCMC_params_n=-4} for each of the e-fold numbers $N_{e}$, seed BH masses $\mBHseed$, and predicted observed BH masses $\mBHtheory$ for the $n = -4$ power law.}
\label{fig:derived_params_n=-4}
\end{figure}

Specifically, we sought, via statistical methods, to fit two samples, consisting of 3 and 8 quasars, respectively. Unlike previous early studies, we explored in detail the case where the uSIDM cross section is velocity-dependent. Our main result is that there exist two quasi-degenerate solutions for certain velocity dependent cross sections in the 3 quasar sample  (Fig.~\ref{fig:MCMC_params}), but with the larger 8 quasar sample the degeneracy generally is removed, and only one ``solution'' remains (Fig.~\ref{fig:MCMC_params_8_quasar}). Thus, with more data, the underlying degeneracies within the uSIDM model can be resolved. In all cases, we find that the uSIDM scenario successfully produces seeds for early-forming SMBH. Despite this success, we find that the $n = 0$ and $n = 1$ power laws predict unphysically large dark matter halos and are therefore ruled out. But, we do find that the $n = -2$ and $n = -4$ power laws can explain the SMBH masses. Excitingly, the $n = -4$ model corresponds to interaction cross sections behaving like Rutherford's cross section, i.e. with a $1/v^4$ velocity dependence, typical of dark sector interactions mediated by a light, or massless ``dark photon''.

\begin{figure}[h]
\includegraphics[scale=0.85]{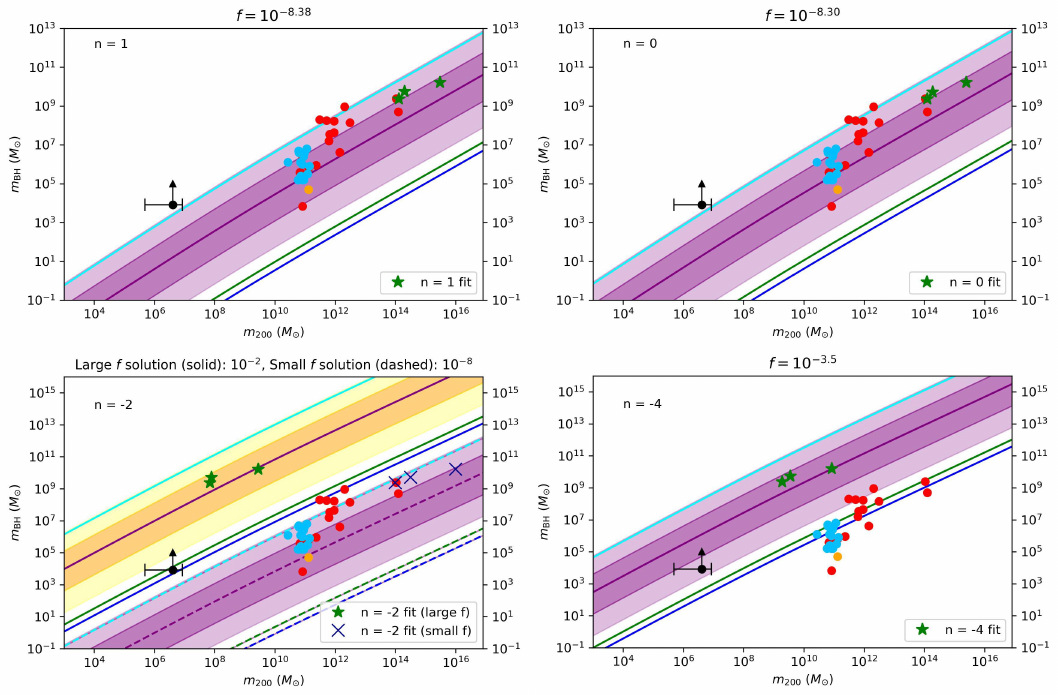}
\caption{The projection of the four velocity dependent cross section fits to the 3 quasar result for the median uSIDM fraction as a function of $m_{200}$ and the number of e-folds, $N_{e}$. The green stars (and blue crosses for $ n = -2$) are the median values of the model fits; there is one red point that overlaps a green star in the $n = 1$ and $n = 0$ sections. The different colored lines have different values for $N_{e}$: blue is 1, green is 2, purple is 10, and light blue is 15. We also plot the bounds for $\omega-$Centauri \cite{DSouza_2013,Haberle_2024} (black dot) as well as other BH's from \cite{Neumayer_2020} (red dots), \cite{Reines_2013} (light blue dots), \cite{Baldassare_2015} (orange dot). We have also plotted the 1$\sigma$(dark purple) and 2$\sigma$(light purple) bands around the median uSIDM fraction $N_{e} = 10$ line (except for the large $f$ solution, for clarity we instead plot orange and yellow for the 1$\sigma$ and 2$\sigma$ bands respectively).}
\label{fig:model_projection}
\end{figure}

\begin{figure}[h]
\centering
\includegraphics[scale=0.55]{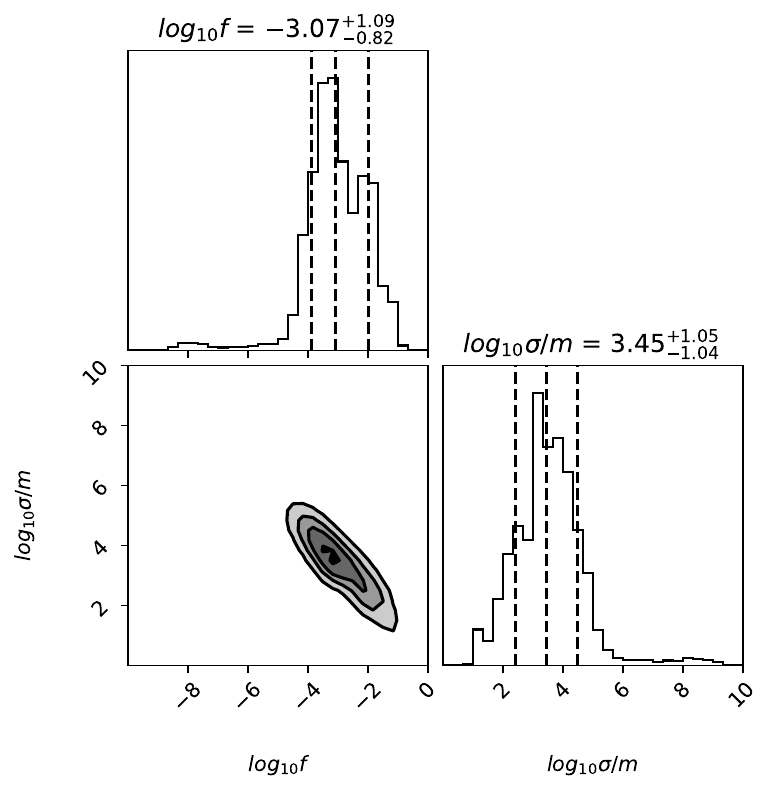}
\includegraphics[scale=0.55]{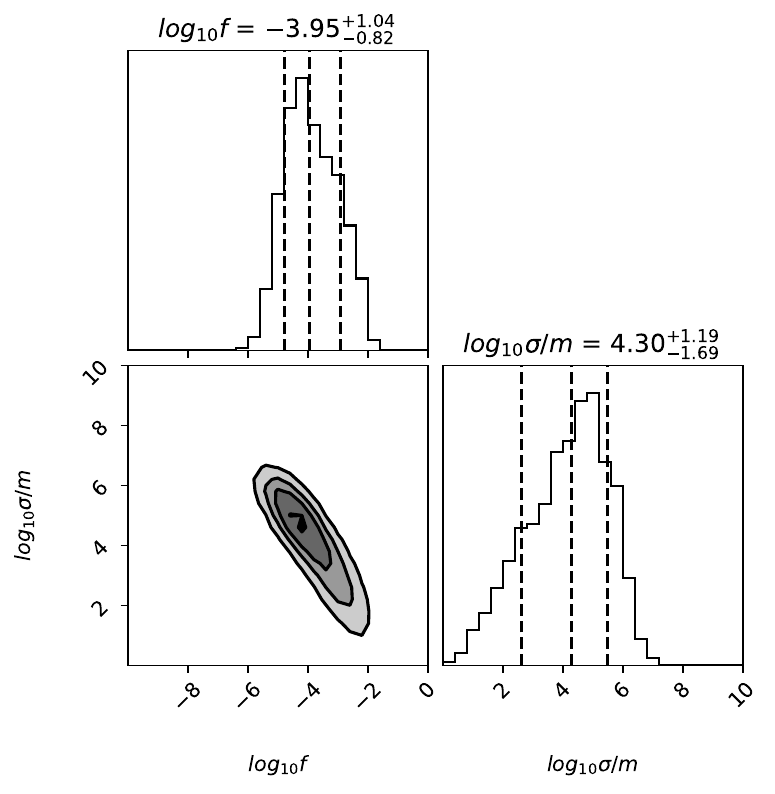}
\caption{The MCMC corner plot for the 8 quasar sample for the uSIDM fraction ($f$) and uSIDM cross section ($\cross$ in $\cmg$) for the $n = -2$ velocity power law model (left) and the $n= -4$ velocity power law model (right). Notice that the dual mode solution for $n = -2$ is effectively converged only to one mode. A very small set of parameter space exists for a second mode (small $f$ and large $\cross$) but it is highly suppressed.}
\label{fig:MCMC_params_8_quasar}
\end{figure}

A similarly exciting prediction of the scenario under consideration is that SMBH seeds should be ubiquitous, in a self-similar way, to dark matter halos of any size. As a result, we predict the existence of intermediate-mass black holes in small galaxies; this is especially exciting given the circumstantial evidence for IMBHs for instance in $\omega$-Centauri (Fig.~\ref{fig:model_projection}), assuming the latter is a dwarf galaxy and not a globular cluster.

Several directions exist that could extend the present study: in order to improve the constraints on the uSIDM parameter space, one could find quasars with both measured dark matter halo masses and SMBH masses. This would reduce the complexity of the model while also narrowing the region of uSIDM parameters allowed. Similarly, with the advent of JWST, more extremely high-redshift quasars, $z\gtrsim10$ (see e.g., \cite{Bogdan2024}) are being discovered; combining such measurements with the similarly growing population of quasars with redshifts $z \approx 6-7$  would allow us to place more stringent constraints on formation/accretion time scales for the seeds. An additional critical input would be  an SIDM-specific concentration-mass relation that one could incorporate into the model, particularly if applicable and available at large redshifts.

\acknowledgments

This work is partly supported by the U.S.\ Department of Energy grant number de-sc0010107 (SP). 

\clearpage
\newpage

\clearpage\newpage


\bibliographystyle{JHEP}

\bibliography{biblio.bib}
\end{document}